\documentclass[sn-mathphys,Numbered]{sn-jnl}
\usepackage{graphicx}%
\usepackage{multirow}%
\usepackage{amsmath,amssymb,amsfonts}%
\usepackage{amsthm}%
\usepackage{mathrsfs}%
\usepackage[title]{appendix}%
\usepackage{xcolor}%
\usepackage{textcomp}%
\usepackage{manyfoot}%
\usepackage{booktabs}%
\usepackage{algorithm}%
\usepackage{algorithmicx}%
\usepackage{listings}%
\usepackage[makeroom]{cancel}
\usepackage{tabularx,subcaption,fancyhdr,mathtools}
\usepackage{hyperref}
\usepackage{multirow}
\usepackage{booktabs}
\usepackage{comment}
\usepackage[T1]{fontenc} % if needed
\usepackage{booktabs}
\usepackage{tabularx}
\DeclareUnicodeCharacter{2212}{-}
\theoremstyle{thmstyleone}%

\theoremstyle{thmstyletwo}%

\theoremstyle{thmstylethree}%

\def \lspone{\widetilde\chi_1^0}
\def \mlspone{m_{\lspone}}
\def \lsptwo{\widetilde\chi_2^0}

\def\chonepm{\widetilde{\chi}_1^{\pm}}
\def\chonemp{\widetilde{\chi}_1^{\mp}}

\def\slep{\widetilde{l}_{L}^{\prime}}
\def\mslep{m_{\slep}}
\def\snu{\widetilde{\nu}}

\def \met{\rm E{\!\!\!/}_T}

\def \lum{\mathcal{L}}
\def \ifb{\rm {fb}^{-1}}

\begin{document}

%\title[Article Title]{Slepton searches in the RPV SUSY scenarios with nonzero trilinear coupling at the HL-LHC and HE-LHC}
\title{\textbf{Slepton searches in the trilinear RPV SUSY scenarios at the HL-LHC and HE-LHC}}

\author[1]{\fnm{Arghya} \sur{Choudhury}}\email{arghya@iitp.ac.in}

\author[1]{\fnm{Arpita} \sur{Mondal}}\email{arpita\_1921ph15@iitp.ac.in}
%\equalcont{These authors contributed equally to this work.}

\author[2]{\fnm{Subhadeep} \sur{Mondal}}\email{subhadeep.mondal@bennett.edu.in}
%\equalcont{These authors contributed equally to this work.}
\author[1]{\fnm{Subhadeep} \sur{Sarkar}}\email{subhadeep\_1921ph21@iitp.ac.in}

\affil[1]{\orgdiv{Department of Physics}, \orgname{Indian Institute of Technology Patna}, \orgaddress{ \city{Patna}, \postcode{801106},  \state{Bihar}, \country{India}}}

\affil[2]{\orgdiv{Department of Physics, SEAS}, \orgname{Bennett University}, \orgaddress{\street{Greater Noida}, \city{Uttar Pradesh}, \postcode{201310}, \country{India}}}

%\affil[3]{\orgdiv{Department}, \orgname{Organization}, \orgaddress{\street{Street}, \city{City}, \postcode{610101}, \state{State}, \country{Country}}}

\abstract{In this work we have studied a multi-lepton final state arising from sneutrino and left-handed slepton production at the high luminosity and high energy LHC in the context of R-parity violating supersymmetry when only the lepton number violating $\lambda_{121}$ and/or $\lambda_{122}$ couplings are non-zero. We have taken into account both pair production and associated production of the three generations of left-handed sleptons and sneutrinos, which are assumed to be mass degenerate. The lightest supersymmetric particle is assumed to be bino and it decays via the R-parity violating couplings into light leptons and neutrinos. Our final state has a large lepton multiplicity, $N_{l}\geq 4~(l=e,~\mu)$. We perform both cut-based and machine learning based analyses for comparison. We present our results in the bino-slepton/sneutrino mass plane in terms of exclusion and discovery reach at the LHC. Following our analysis, the slepton mass can be discovered upto $\sim$ 1.54 TeV and excluded upto $\sim$ 1.87 TeV at the high luminosity LHC while these ranges go upto $\sim$ 2.46 TeV and $\sim$ 3.06 TeV respectively at the high energy LHC.}

\maketitle

\section{Introduction}
\label{sec:intro}

The LHC collaborations have meticulously looked for the signal of beyond Standard Model (BSM) physics using Run-I and Run-II data and will do the same with ongoing Run-III operation. Supersymmetry \cite{Drees:2004jm, Haber:2017aci, Martin:1997ns, Nilles:1983ge} is still the most popular and promising BSM scenario that solves  various shortcomings of the standard model (SM) - e.g., the gauge hierarchy problem \cite{SUSSKIND1984181,PhysRevD.14.1667}, observed dark matter (DM) relic density of the universe \cite{Planck:2018vyg}, muon (g-2) anomaly \cite{Muong-2:2021ojo} etc. As there are still no hints of new physics signal from the LHC Run-I and Run-II data, the lower bound on strongly interacting colored sparticle masses have reached up to $\mathcal{O}$(2-2.5) TeV \cite{cms_susy,atlas_susy}. On the other hand, the bounds on the electroweak (EW) sector SUSY particles are much weaker \cite{cms_susy,atlas_susy}. The R-parity conserving (RPC) minimal supersymmetric standard model (MSSM)~\cite{Drees:2004jm, Haber:2017aci, Martin:1997ns, Nilles:1983ge,Vempati:2012np} provides a stable weakly interacting massive particle (WIMP) which can be a natural DM candidate \cite{Bertone:2004pz,Baer:2008uu,Roszkowski:2004jc, Jungman:1995df}  and the most popular choice 
is the lightest neutralino ($\lspone$) in the form of lightest SUSY particle (LSP). In the RPC scenarios with light EW sectors there are a large number of phenomenological analyses which have addressed the implication of LHC results along with muon (g-2) anomaly and observed DM relic density data \cite{Barman:2022jdg, He:2023lgi, Chakraborti:2017dpu, Chowdhury:2016qnz, Bhattacharyya:2011se, Choudhury:2012tc, Choudhury:2013jpa, Chakraborti:2021bmv, Choudhury:2016lku, Dutta:2015exw, Dutta:2017jpe}. 

The RPC MSSM is more extensively studied in literature due to the DM candidate in the form of the LSP and for that, we need to incorporate the R-parity conservation by hand. If R-parity violating (RPV) terms are allowed, the superpotential looks like \cite{Dreiner:1997uz,Barbier:2004ez,Banks:1995by}
\begin{equation} \label{eq:rpv_potential}
W_{\cancel{R}_p} = \mu_i\hat{H}_u.\hat{L}_i + \frac{1}{2}\lambda_{ijk}\hat{L}_i.\hat{L}_j\hat{e}_k^c + \frac{1}{2}\lambda^\prime_{ijk}\hat{L}_i.\hat{Q}_j\hat{d}_k^c + \frac{1}{2}\lambda^{\prime\prime}_{ijk}\hat{u}_i^c\hat{d}_j^c\hat{d}_k^c
\end{equation}
where the first three terms are the lepton number violating terms and the last term violates the baryon number. Here $\hat{H}_u$ is the up-type Higgs supermultiplet and $\hat{L}$ ($\hat{e}$) refers to the left-handed lepton doublet (right-handed singlet) supermultiplet. Similarly, $\hat{Q}$ ($\hat{u}$) corresponds to the up-type left (right)-handed doublet (singlet) quark supermultiplet. $\hat{d}$ represents the right-handed down-type quark supermultiplet. In this work we only consider the non-zero $\lambda$ couplings\footnote{These couplings contribute to light neutrino masses and mixings at one loop level \cite{Grossman:2003gq, Barbier:2004ez} and to muon (g-2) \cite{Kim:2001se, Chakraborty:2015bsk}.} which have distinctive collider signatures compared to RPC scenarios. In the RPC scenario, the stable LSP leads to a large amount of missing energy $\met$ in the final states while the LSP decays to multilepton final states for  $\lambda_{ijk} \ne 0$ scenarios.  Depending on the choices of LSP and non-zero RPV couplings, one obtains various novel final states \cite{Dreiner:2023bvs,Mitsou:2015kpa,Bardhan:2016gui,Bhattacherjee:2013tha,Bhattacherjee:2013gr,Dercks:2018eua}. For different choices 
of  $\lambda_{ijk}$ couplings, the LHC collaborations have already derived limits on chargino and slepton masses \cite{ATLAS:2021yyr} using LHC Run-II data and it will be interesting to study the sensitivity of the EW sparticle searches at the 14 TeV high luminosity LHC (HL-LHC) and the proposed high energy (27 TeV) upgrade of the LHC (HE-LHC). It may be noted that the SUSY parameter space containing one or more lighter EW sparticles like charginos, sneutrinos, smuons or neutralinos are very much consistent with the recent measurement of muon magnetic moment at Fermilab \cite{Muong-2:2006rrc, Muong-2:2021ojo, Muong-2:2023cdq}. The additional contributions from SUSY mainly come from the chargino-sneutrino loop and smuon-neutralino loop and there could be even some additional contribution in the RPV scenarios depending on the couplings. A few phenomenological analyses with RPC and RPV scenarios in the context of muon (g-2) anomaly may be seen in Refs. 
\cite{Chakraborti:2022vds, Choudhury:2023lbp, He:2023lgi, Chakraborti:2021bmv, Chakraborti:2015mra, Chakraborti:2014gea, Baer:2021aax, Endo:2021zal, Choudhury:2017acn, Kowalska:2015zja, Hundi:2011si, Altmannshofer:2020axr, Zheng:2021wnu, Chakraborty:2015bsk,Zheng:2022ssr}. 

In a recent work~\cite{Choudhury:2023eje} the search prospect of gaugino sector at the HL-LHC and HE-LHC with $\lum = 3000~\ifb$ is presented using the direct $\chonepm\chonemp$ and $\chonepm\lsptwo$ production for scenarios with non-zero $\lambda_{ijk}$ couplings\footnote{For electroweakino searches in the context of $UDD$ couplings at the HL-LHC please refer to~\cite{Barman:2020azo}.}. In this work we extend the similar analysis using the direct production of mass degenerate L-type sleptons of all three generations. First we look for the results using traditional cut-and-count based analysis and then look for the improvement of sensitivity using a machine learning (ML) algorithm. For the ML analysis we will adopt the boosted decision tree (BDT) \cite{Cornell:2021gut,Coadou:2022nsh} algorithm. 

In Sec.~\ref{sec:model} we discuss the model framework along with the possibility of different final states arising from distinct choice of RPV couplings. 
In Sec.~\ref{sec:collider}, we first present the projected exclusion limits in the slepton-LSP mass plane using a traditional cut-and-count analysis followed by a ML based algorithm at the HL-LHC. We extend this multilepton analysis for the HE-LHC also. We conclude our results in Sec.~\ref{sec:conclusion}.

%%%%%%%%%%%%%%%%%%%%%%%%%%%%%%%
\section{Model Definition}
\label{sec:model}
%%%%%%%%%%%%%%%%%%%%%%%%%%%%%%
The pair production cross-sections for L-type charged slepton is roughly $\sim 3$ times larger than the R-type charged slepton pair production \cite{Fuks:2013lya, Fiaschi:2018xdm, Fiaschi:2023tkq, Bozzi:2007qr, Fuks:2013vua, Beenakker:1999xh}. In the context of RPC slepton searches, the most popular simplified models consist of both L and R-type charged slepton of the first two generations. On the other hand, the sneutrino pair productions or charged slepton-sneutrino productions contribute to the multilepton final states in the RPV SUSY scenarios with nonzero $\lambda_{ijk}$ couplings. In this work, we consider a simplified model where all three generations L-type charged sleptons and sneutrinos are mass degenerate (R-type sleptons are assumed to be lying beyond the reach of the LHC) and the sleptons are produced via $pp \rightarrow \slep\slep, \snu\snu$ and $\slep\tilde{\nu}$ channels, where $l^{\prime} \equiv e, \mu, \tau$ The corresponding Feynman diagrams for these productions are shown in Fig.~\ref{fig:rpv_decay}.
%%%%%%%%%%%%%%%%%%%%%%%%%%%%%%%%%%%
\begin{figure}[h]
\centering
\includegraphics[width=0.3\linewidth]{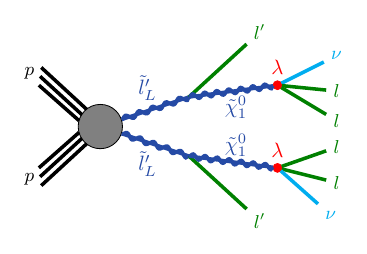}
\includegraphics[width=0.3\linewidth]{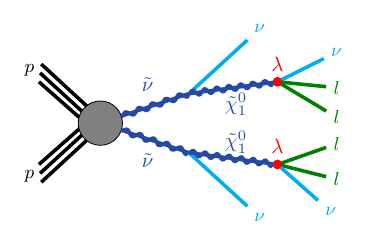}
\includegraphics[width=0.3\linewidth]{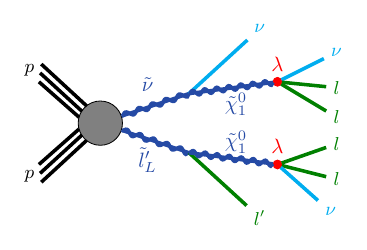}
\caption{Diagrams for decays of $\slep$, $\tilde{\nu}$ to $\lspone$ and $\lspone$ decays to leptons via $\lambda$ coupling. Here $l^{\prime} = e, \mu , \tau$ and $l = e, \mu$}
\label{fig:rpv_decay}
\end{figure} 
%%%%%%%%%%%%%%%%%%%%%%%%%%%%%%%%%%%%%%
Here the sleptons decay to lepton and bino type LSP ($\lspone$) and via the $\lambda_{ijk}$ couplings the LSP decays into $l_k^{\prime\pm}l_{i/j}^{\prime\mp}\nu_{j/i}$, where $l^{\prime} = e, \mu, \tau$. Thus one obtains maximally enriched 
leptonic final states in scenarios with non-zero $\lambda_{121}$ and/or $\lambda_{122}$ couplings where the LSP pair in the final state always gives rise to $4l$ ($l = e, \mu$) + $\met$ topology\footnote{For example $\lspone$ decays to $e e \nu_\mu$ and  $e \mu  \nu_e$  with 50\% branching ratios each for single non-zero values of $\lambda_{121}$ coupling}. Depending upon the production modes 
($\slep\tilde{\nu}_L$ and $\slep\slep$) one or two more leptons may arise in the final state. In this work, we will focus on non-zero $\lambda_{12k}$ ($k~\epsilon~1,2$) scenarios and consider the final states consisting of at least four leptons $N_l \geq 4$, where $l \equiv e, \mu$. It may be noted that 9 non-zero $\lambda_{ijk}$ couplings lead to different charged lepton configurations and the LSP pair in the final states leads to three more different scenarios where the leptonic ($l = e, \mu$) branching ratios get reduced and 
the collider limits or sensitivity become weaker~\footnote{Before concluding our 
results we will also briefly comments on the sensitivity in such scenarios.}. For more details on the 
LSP decay modes and charged lepton configurations for various scenarios see Sec.~2 of Ref.~\cite{Choudhury:2023eje}.   
  
%%%%%%%%%%%%%%%%%%%%%%%%%%%%%
\section{Collider analysis}
\label{sec:collider}
%%%%%%%%%%%%%%%%%%%%%%%%%%%%%
As discussed in the previous section, we consider a simplified model with L-type mass degenerate charged sleptons and sneutrinos (all three generations) and look for the projection reach at the HL-LHC and HE-LHC using a final state with at least 4 leptons ($N_{l} \geq 4$). For this final state, dominant SM backgrounds are $ZZ + jets$, $WWZ + jets$ and $t\bar{t}Z + jets$ and we also compute sub-dominant processes like $WZZ + jets$, $ZZZ + jets$, h production via $ggF$, $hjj$, $Wh + jets$, $Zh + jets$. All these  SM backgrounds and the SUSY signals have been generated using \texttt{MadGraph5-aMC@NLO} \cite{Alwall:2014hca} and \texttt{Pythia-6.4} \cite{Sjostrand:2006za} at the leading order (LO) parton level respectively. The cross-sections for the SM backgrounds are taken from Table 11 and Table 13 of Ref.~\cite{Choudhury:2023eje}. The NLO+NLL level cross-sections for SUSY signals have been calculated using \texttt{Resummino-3.1.1} \cite{Fuks:2013lya, Fuks:2013vua, Fiaschi:2018xdm, Fiaschi:2023tkq, Bozzi:2007qr, Beenakker:1999xh}. For fast detector simulation we have used \texttt{DELPHES 3}  (version-3.5.0) platform \cite{deFavereau:2013fsa}. Jets reconstructions have been done using anti-$k_t$ algorithm \cite{Cacciari:2008gp} algorithm with jet radius $R = 0.4$ , $p_T > 20$ and $|\eta| < 2.8$. Leptons are reconstructed with $p_T > 7~(5)$ and $|\eta| < 2.47~(2.7)$ cuts from electron (muon) candidates after isolation. The track and calorimeter isolation, $b-$tagging efficiency, generation level cuts, jet matching, etc. are done in a similar way as prescribed in Sec.3 of Ref.\cite{Choudhury:2023eje}.

\subsection{Prospect at the HL-LHC: cut-based vs. ML analysis}
\label{sec:14_cut}

In this section, we present the prospect of slepton pair productions at the high luminosity LHC (HL-LHC) with $N_l \geq 4$ final state at the center of mass energy $\sqrt{s} = 14$  TeV and $\mathcal{L} = 3000$ fb$^{-1}$. First, we will present the conventional cut-and-count analysis and then we will look for the improvement using a boosted decision tree (BDT) based machine learning algorithm.
%%%%%%%%%%%%%%%%%%%%%%%%%%%%%%%%%%%%%%%
\begin{figure}[!htb]
\centering
\includegraphics[width=0.45\linewidth]{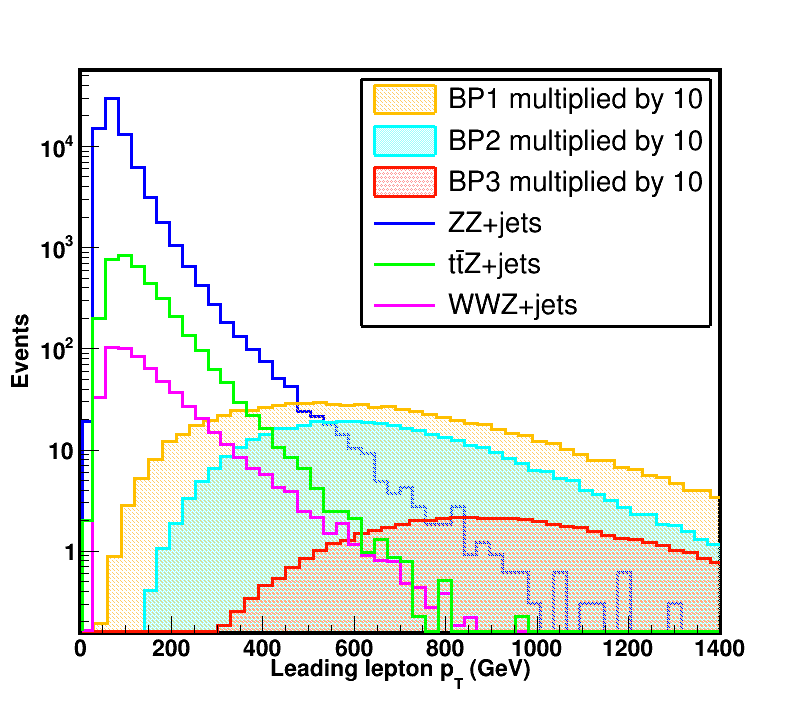}
\includegraphics[width=0.45\linewidth]{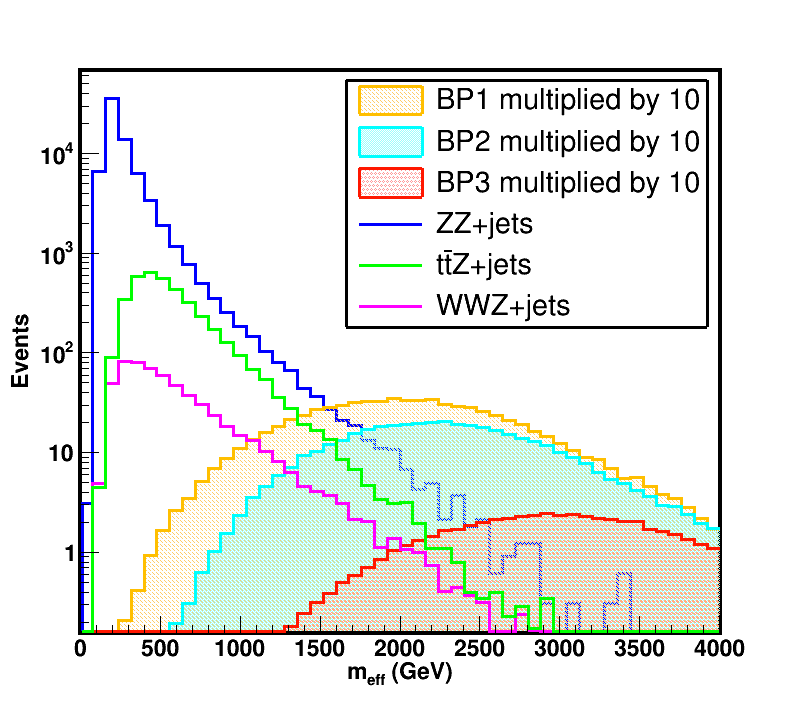}
\caption{Distributions of transverse momentum of leading lepton - $p_T^{l_1}$ (left) and effective mass - $m_{eff}$ (right) at the HL-LHC 
   ($\sqrt{s}=14$ TeV with $\mathcal{L}=3000$ $fb^{-1}$) are shown here.  
The blue, green, and magenta color solid lines represent the most dominant $ZZ + jets$, $t\bar{t}Z+jets$ and $WWZ+jets$ backgrounds. 
Yellow, cyan, and red filled regions correspond to the benchmark points - \texttt{BP1}, \texttt{BP2}, and \texttt{BP3} respectively with a multiplication factor of 10.}
\label{fig:lep_meff14}
\end{figure}
%%%%%%%%%%%%%%%%%%%%%%%%%%%%%%%%%%%%%%

The ATLAS collaboration has already excluded the slepton mass up to 1.2 TeV using RUN-II LHC data~\cite{ATLAS:2021yyr} for non-zero $\lambda_{12k}$ ($k~\epsilon~1,2$) couplings. Following the ATLAS analysis~\cite{ATLAS:2021yyr} and similar to our previous work~\cite{Choudhury:2023eje}, we optimize the effective mass ($m_{eff}$) variable for cut-based analysis to maximize the signal significance and define the signal regions. To demonstrate the results we consider three signal benchmark points: 
- \texttt{BP1}: $\mslep = 1300$ GeV, $\mlspone = 250$ GeV; \texttt{BP2}: $\mslep = 1450$ GeV, $\mlspone = 800$ GeV; \texttt{BP3}: $\mslep = 1800$ GeV, $\mlspone = 1750$ GeV. The mass difference between sleptons and the LSP is large, intermediate, and small for  \texttt{BP1}, \texttt{BP2}, and \texttt{BP3} respectively. In Fig.~\ref{fig:lep_meff14}, we depict the distribution of leading lepton transverse momentum ($p_T^{l_1}$) and effective mass ($m_{eff}$)\footnote{The effective mass is defined as $m_{eff}=\sum_i p_T^{l_i} + \sum_i p_T^{j_i} + \met$.} for the three benchmark points along with the three main SM backgrounds ($ZZ + jets$, $WWZ + jets$, $t\bar{t}Z + jets$). The yellow, cyan, and red filled regions correspond to \texttt{BP1}, \texttt{BP2}, and \texttt{BP3} respectively, while the blue, green, and magenta colored lines illustrate the $ZZ+ jets$,  $t\bar{t}Z + jets$ and $WWZ + jets$ background respectively. 
Fig.~\ref{fig:lep_meff14} suggests that the typical cuts like $p_T^{l_1} > 100$ GeV and a large $m_{eff}$ will reduce the backgrounds significantly without affecting the signal events too much. Following cut optimization we observe that the $N_{l} \geq$ 4 + $p_T^{l_1} > 100$ + $Z$ veto\footnote{The event with same flavor opposite sign leptons pair with invariant mass range $81.2 < m_{SFOS} < 101.2$ GeV are excluded.} + $b$ veto cut along with large $m_{eff}$ provides the maximum signal significance. We define 
the two signal region as : \texttt{SR-A} with $m_{eff} > 900$ GeV and \texttt{SR-B} with $m_{eff} > 1500$ GeV. 

%%%%%%%%%%%%%%%%%%%%%%%%%%%%%%%%%%%%%%%%%%%%%%%%%%%
\begin{table}[!htb]
\begin{tabular}{||c|c|c|c|c|c||}
\hline
\hline
  &  &  &  & \multicolumn{2}{c||}{\underline{Signal Region}} \\ %\cline{5-6}
   & $N_l\ge 4  ~(l=e,\mu)$ & & & & \\
   &  +  &  &  & SR-A  & SR-B \\
    & $p_T^{l_1} >$ 100 GeV  & Z veto& b veto& ($m_{eff}>900$ GeV)& ($m_{eff}>1500$ GeV) \\ 
 \hline 
 \hline
\texttt{BP1} (1300,250)  & 74.45 & 65.28 & 60.87 & 59.16 & 49.10 \\ 
$\sigma_{NLO+NLL}^{14}=0.0381\rm~{fb}$ & & & & & \\ \hline
\texttt{BP2} (1450,800)  & 42.20 & 40.99 & 38.01 & 37.80 & 34.01 \\ 
$\sigma_{NLO+NLL}^{14}=0.0196\rm~{fb}$ & & & & & \\ \hline
\texttt{BP3} (1800,1750) & 5.86 & 5.56 & 5.20 & 5.19 & 5.10\\ 
$\sigma_{NLO+NLL}^{14}=0.0029\rm~{fb}$ & & & & & \\ \hline
\hline
Total & & & & & \\
 background & 22124.17 & 382.82 & 221.92 & 20.19 & 3.498 \\
\hline
\hline
\multicolumn{2}{||c|}{\multirow{3}{*}{\shortstack{Signal Significance $\sigma_{ss}^0$ \\ ( $\sigma_{ss}^{5}$, Sys. Unc.~$\epsilon$=5\%)}}}   &\multicolumn{2}{c|}{\texttt{BP1}} & 6.64 (6.07) & 6.77 (6.36) \\  %\cline{3-6}
\multicolumn{2}{||c|}{}& \multicolumn{2}{c|}{\texttt{BP2}} & 4.96 (4.64) & 5.55 (5.31)\\ %\cline{3-6} 
\multicolumn{2}{||c|}{}& \multicolumn{2}{c|}{\texttt{BP3}}  & 1.03 (0.998) & 1.74 (1.72) \\ %\cline{3-6}
\hline
\hline
\end{tabular}
\caption{ Production cross-sections, selection cuts, and the corresponding yields at the HL-LHC for the three signal benchmark points are presented. The total SM background yields are also shown. The last three rows represent the signal significance  ($\sigma_{ss}^0$) 
without any systematic uncertainty and $\sigma_{ss}^{5}$  (with Sys. Unc. $\epsilon$ = 5\%) for the benchmark points.}
\label{tab:cut_flow_14}
\end{table}
%%%%%%%%%%%%%%%%%%%%%%%%%%%%%%%%%%%%%%%%%%%%%%%%%%%%%%%%%%%%

In Table~\ref{tab:cut_flow_14}, we have summarized the production cross-section of the signal benchmark points, the corresponding yields of signal events, and the total background yield after the selection cuts. The signal significance without systematic uncertainty ($\sigma_{ss}^0$) and with 5\% uncertainty ($\sigma_{ss}^{5}$) are also presented in the last three rows. To estimate the signal significance we have used the relation $\sigma_{ss}^{\epsilon} = S/\sqrt{S+B+((S+B)\epsilon)^2}$ where S, B and $\epsilon$ correspond to the signal yield, background yield and systematic uncertainty respectively. Due to the large $m_{eff}$ cut, the \texttt{SR-B} signal regions are more effective than \texttt{SR-A} in the parameter space where slepton masses are relatively higher. The signal significance corresponding to  \texttt{BP1}, \texttt{BP2} and \texttt{BP3}  for \texttt{SR-B} signal region are 6.77 (6.35), 5.55 (5.31), 1.74 (1.72) respectively for $\epsilon$  = 0\% (5\%). The effect of including 5\% uncertainty is not significant and it is observed that the signal significance reduces by 1-6\%. We illustrate the projected discovery ($5\sigma$) and exclusion (2$\sigma$) reach at the HL-LHC using cut-based methods in $\mslep-\mlspone$ mass plane in Fig.~\ref{fig:reach_14tev_ml}. The light-blue filled region denotes the $2\sigma$ projection and the blue dotted line corresponds to the projected 5$\sigma$ discovery reach. The projected exclusion (discovery)  curve reaches up to $\sim$ 1.75 (1.49) TeV on slepton masses.

Next, we perform a ML-based analysis for the improvement of signal significance using an Extreme Gradient Boosted decision tree algorithm through {\tt XGBoost} machine learning toolkit \cite{Chen:2016btl}. A set of 18 `features' (kinematic variables) are constructed to perform our analysis which are transverse momenta of leading and subleading lepton ($p_T^{l_1}$ and $p_T^{l_2}$), $\Delta R_{l_il_j}$ between first four leading leptons\footnote{$\Delta R$ is defined as $\Delta R= \sqrt{(\Delta \eta)^2 + (\Delta \phi)^2}$} (6 variables), the difference in azimuthal angle $\Delta \phi_{l_i \met}$ between first four leading leptons and $\met$ (4 variables),  number of b-tagged ($N_b$), non-b-tagged jets ($N_j$),   missing transverse energy ($\met$), effective mass ($m_{eff}$), number of SFOS pair ($N_{SFOS}$) and number of reconstructed $Z$ ($N_Z$). We have considered $N_l\geq4$ ($l\equiv e, \mu$) events for SUSY signal and SM backgrounds.  
 In this process, 80\% of the data set is considered for training and the rest for testing using {\tt multi:softprob} objective function to achieve the multiclass classification. We choose the hyperparameters as follows: \texttt{number of trees} = 500, \texttt{maximum depth} = 10, and \texttt{learning rate} = 0.03 for optimal outcome. Finally, we obtain the predicted probability score of different classes for every event and applying a threshold on this score we estimate the maximum significance. 
 
 %%%%%%%%%%%%%%%%%%%%%%%%%%%%%%%%%%%%%
\begin{figure}[h]
\centering
\includegraphics[width=0.45\linewidth]{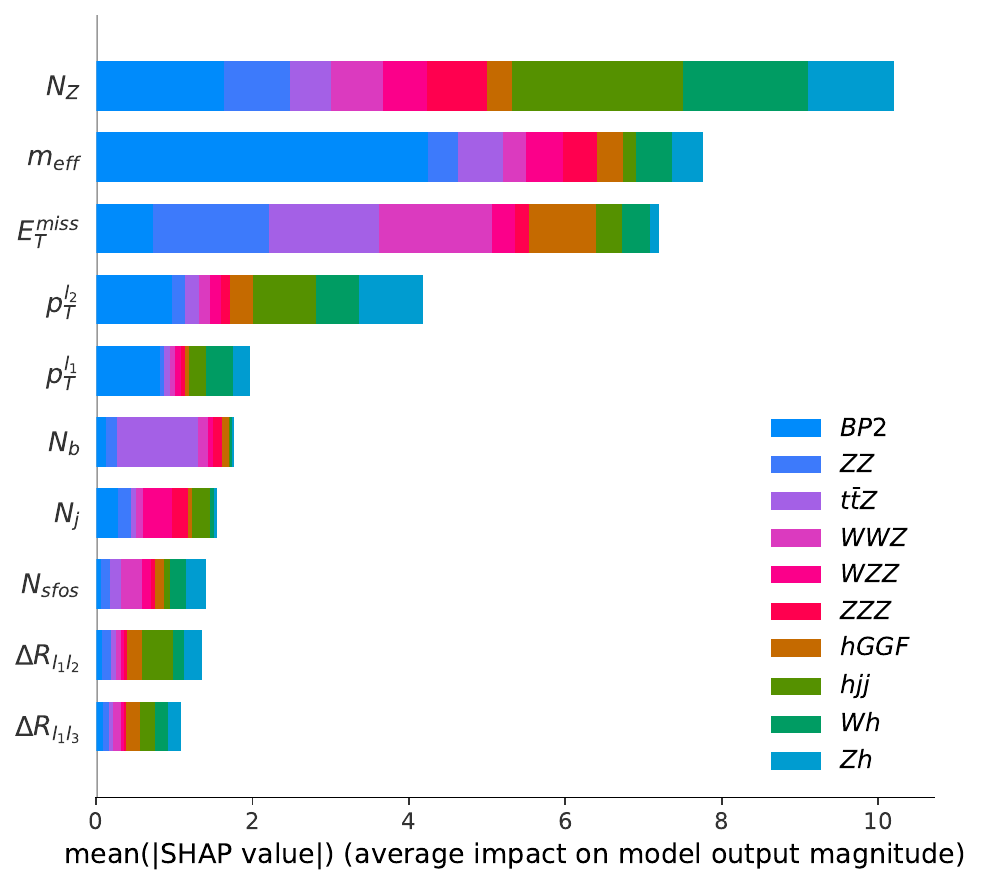}
\includegraphics[width=0.45\linewidth]{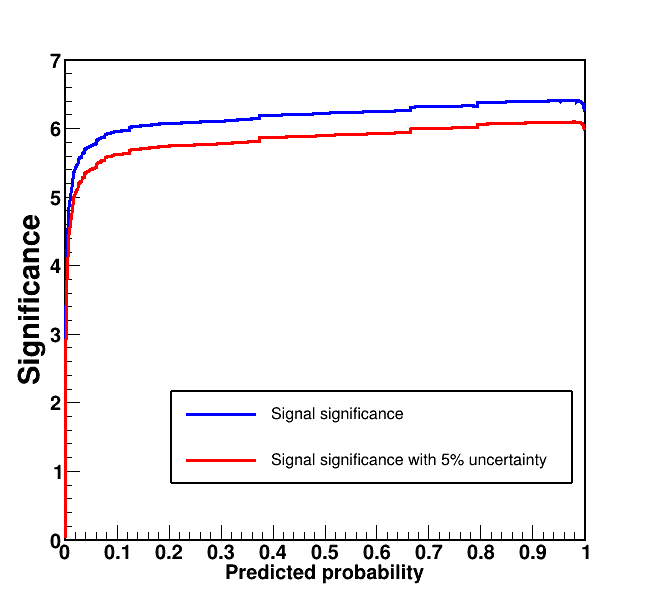}
\caption{Left: Shapley feature importance plot for the top ten variables for benchmark point BP2 and the SM backgrounds. Right: The signal significance without any systematic uncertainty and with 5\% uncertainty are displayed via blue and red lines as a function of predicted probability.}
\label{fig:shap_scoreBP2}
\end{figure}
%%%%%%%%%%%%%%%%%%%%%%%%%%%%%%%%%%%%%

To find out the effectiveness of each feature and their ranking we have estimated the Shapley values using SHAP (SHapley Additive exPlanation) \cite{DBLP:journals/corr/LundbergL17,Shapley+1953+307+318} toolkit. We plot the mean of absolute values of the top 10 kinematic variables in Fig.~\ref{fig:shap_scoreBP2} (left panel) for a particular signal benchmark point, \texttt{BP2} ($\mslep=1450$ GeV, $\mlspone=800$ GeV) and SM backgrounds. 
The top three important variables are $N_Z$, $\met$ and $m_{eff}$. We also depict the effect of probability scores on the signal significance for \texttt{BP2} with 0 $\%$ (blue line) and 5 $\%$ (red line) systematic uncertainty in the right panel of Fig.~\ref{fig:shap_scoreBP2}. It is evident that the signal significance saturates around probability score $\sim$ 0.9-0.96. 

The signal yields, the total background yields, the signal significances for the benchmark points, and the gain in significance from cut-based analysis are demonstrated in Table~\ref{tab:significance_14_ml} for probability score values 0.90 and 0.96 . The numbers in the brackets are the results corresponding to significance and gain with 5\% systematic uncertainty. Around 15-38\% gain in signal significance is achieved by implementing ML algorithms as compared to the cut-based method.  
%%%%%%%%%%%%%%%%%%%%%%%%%%%%%%%%%%%%%%%%%%%%%%
\begin{table}[!htb]
%\small
\begin{tabular}{|c|c|c|c|c|c|}
\hline
Benchmark Points & Probability Score& Signal   & Total   & Signal  &  Gain in $\sigma_{ss}$ \\
 &  &Yield  & Background Yield &  Significance $\sigma_{ss}$ & from  \\
& &  &  &  (Sys Unc. = $5\%$) & Cut-based \\
\hline
\hline
\texttt{BP1} & 0.90  & 72.66 & 3.61  & 8.32 (7.62) & 23\% (20\%)\\
%\cline{2-6}
 & 0.96 & 71.70 & 1.98 & 8.35 (7.68)  & 23\% (21\%) \\
\hline
\hline
\texttt{BP2} & 0.90  & 41.82  & 1.17 & 6.38 (6.06) & 15\% (14\%) \\
%\cline{2-6}
 & 0.96 & 41.66  & 0.87 & 6.39 (6.07) & 15\% (14\%) \\
\hline
\hline
\texttt{BP3} & 0.90  & 5.86  & 0.95 & 2.25 (2.23) & 29\% (30\%)  \\
%\cline{2-6}
 &  0.96 & 5.85  & 0.10 & 2.40 (2.38) & 38\% (38\%)\\
\hline
\hline
\end{tabular}
\caption{Signal yield, total background yield, the signal significance and the gain in significance at the HL-LHC using ML-based algorithm for different probability scores are shown here. In the last two columns, the numbers in the parenthesis correspond to signal significance and gain with systematic uncertainty $\epsilon = 5\%$.}
\label{tab:significance_14_ml}
\end{table}
\label{sec:14_cut}
%%%%%%%%%%%%%%%%%%%%%%%%%%%%%%%%%%%%%%%%%%%%%%

%%%%%%%%%%%%%%%%%%%%%%%%%%%%%%%%%%%%%%%%%
\begin{figure}[!htb]
\begin{center}
\scalebox{0.8}{\input{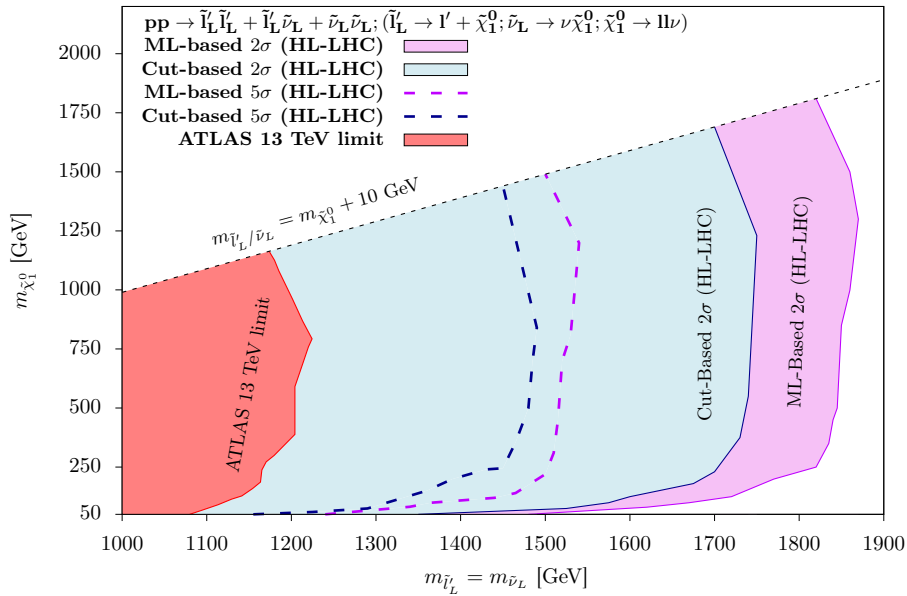}}
\caption {Projected discovery ($5\sigma$) and exclusion ($2\sigma$) reach in the slepton-LSP mass plane at the HL-LHC are presented for cut-based and ML-based analysis. The red regions represent the existing limit obtained by the ATLAS collaboration from Run-II data\cite{ATLAS:2021yyr}. The violet (blue) regions correspond to $2\sigma$ projected reach from the ML-based (cut-and-count) method at the HL-LHC.   The blue and violet dotted line stand for $5\sigma$ projected discovery reach obtained from traditional cut-based method and {\tt XGBoost} respectively.}
\label{fig:reach_14tev_ml}
\end{center}
\end{figure}
%%%%%%%%%%%%%%%%%%%%%%%%%%%%%%%%%%%%%%%%

Finally, we estimate the projected discovery ($5\sigma$) and exclusion ($2\sigma$) reach in $\mslep-\mlspone$ plane at HL-LHC using ML-based methods. To compare the results with cut-and-count analysis we present the $5\sigma$ and  $2\sigma$ projection in Fig.~\ref{fig:reach_14tev_ml} using the violet dotted line 
 and violet-colored filled region respectively. The $5\sigma$ and $2\sigma$ curves extend up to $\sim$ 1.54 TeV and 1.87 TeV for left-handed degenerate slepton masses (all three generations). This provides us an enhancement of approximately 120 GeV improvement in the projected exclusion limits using {\tt XGBoost} compared to the cut-based method.

\subsection{Prospect at the HE-LHC: cut-based vs. ML analysis}
\label{sec:27_cut}

We further extend our analysis for High Energy LHC ($\sqrt{s}=27$ TeV, $\mathcal{L}=3000~\rm{fb^{-1}}$) for direct slepton pair production with $N_l\geq 4$ final state. Similar to Sec.~\ref{sec:14_cut}, we implement the cut-and-count method along with ML-based algorithm for further improvement. To showcase our results and compare the two methods we have chosen three signal benchmark points as :
\texttt{BP1} ($\mslep = 1300$ GeV, $\mlspone = 250$ GeV), \texttt{BP4} ($\mslep = 2000$ GeV, $\mlspone = 1000$ GeV) and \texttt{BP5} ($\mslep = 2750$ GeV, $\mlspone = 2500$ GeV). 

%%%%%%%%%%%%%%%%%%%%%%%%%%%%%%%%%%%%%%%%
\begin{figure}[!htb]
\centering
\includegraphics[width=0.45\linewidth]{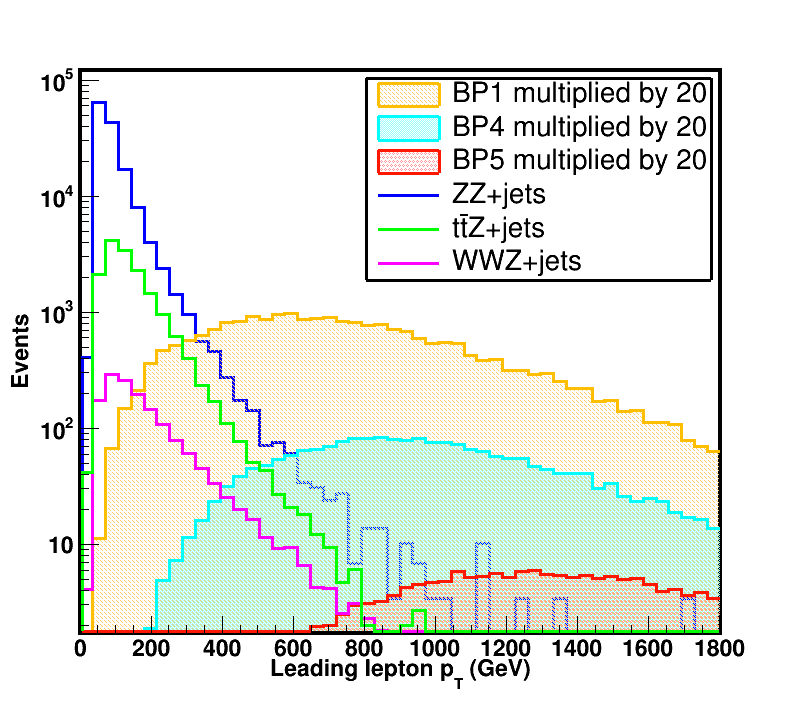}
\includegraphics[width=0.45\linewidth]{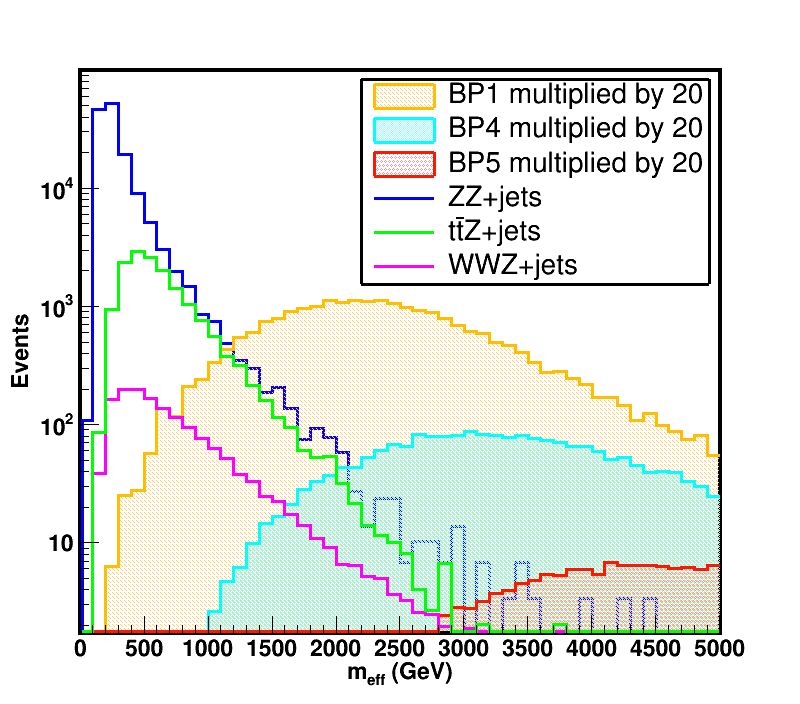}
\caption{Distributions of transverse momentum of leading lepton ($p_T^{l_1}$) and effective mass ($m_{eff}$) at the HE-LHC 
   ($\sqrt{s}=27$ TeV with $\mathcal{L}=3000$ $fb^{-1}$) are shown here. Color conventions for the SM backgrounds are the same as in  Fig.~\ref{fig:lep_meff14}. Yellow, cyan, and red filled regions correspond to the benchmark points - \texttt{BP1}, \texttt{BP4} and \texttt{BP5} respectively (scaled by a factor 20).}
\label{fig:lep_meff27}
\end{figure}
%%%%%%%%%%%%%%%%%%%%%%%%%%%%%%%%%%%%%%%%

%%%%%%%%%%%%%%%%%%%%%%%%%%%%%%%%%%%%%%%%%%%%%
\begin{table}[!htb]
\begin{tabular}{||c|c|c|c|c|c||}
\hline
\hline
  & \multirow{2}{*} {} & \multirow{2}{*}{} & \multirow{2}{*}{} & \multicolumn{2}{c||}{\underline{Signal Region}} \\  %\cline{5-6}
   & \shortstack{$N_l\ge4$   $(l=e,\mu)$ \\+ \\$p_T^{l_1} >$ 150 GeV}  & Z veto & b veto & {\shortstack{SR-C \\($m_{eff}>1500$ GeV)}} & {\shortstack{SR-D \\($m_{eff}>2200$ GeV)}}\\ 
 \hline 
 \hline
\texttt{BP1} (1300,250)  & 1165.63 & 1032.68 & 937.85 & 806.16 & 524.44 \\ 
$\sigma_{NLO+NLL}^{27}=0.628~\rm{fb}$ & & & & & \\ \hline
\texttt{BP4} (2000,1000)  & 110.79 & 109.13 & 98.76 & 97.40 & 87.34 \\ 
$\sigma_{NLO+NLL}^{27}=0.052~\rm{fb}$ & & & & & \\ \hline
\texttt{BP5} (2750,2500) & 10.24 & 10.14 & 9.20 & 9.19 & 9.09 \\ 
$\sigma_{NLO+NLL}^{27}=0.0048~\rm{fb}$ & & & & & \\ \hline
\hline
Total & & & & & \\
 background & 26640.55 & 708.36 & 307.57 & 23.76 & 5.86 \\
\hline
\hline
\multicolumn{2}{||c|}{\multirow{3}{*}{\shortstack{Signal Significance $\sigma_{ss}^0$ \\ ( $\sigma_{ss}^{5}$, Sys. Unc.=5\%)}}}   &\multicolumn{2}{c|}{\texttt{BP1}} & 27.98 (15.96) & 22.77 (14.93)\\  %\cline{3-6}
\multicolumn{2}{||c|}{}& \multicolumn{2}{c|}{\texttt{BP4}} & 8.85 (7.75)& 9.04 (8.14)\\ %\cline{3-6} 
\multicolumn{2}{||c|}{}& \multicolumn{2}{c|}{\texttt{BP5}}  & 1.60 (1.54) & 2.35 (2.31)\\ %\cline{3-6}
\hline
\hline
\end{tabular}
\caption{Cut flow table for signal benchmark points and the total SM backgrounds along with $\sigma_{ss}^{0}$ ($\sigma_{ss}^{5}$)  at the HE-LHC are shown here.}
\label{tab:cut_flow_27}
\end{table}
%%%%%%%%%%%%%%%%%%%%%%%%%%%%%%%%%%%%%%%%%%%%%

Fig.~\ref{fig:lep_meff27} depicts the distributions for transverse momenta of leading lepton ($p_T^{l_1}$) and effective mass ($m_{eff}$) corresponding to these benchmark points and three dominant SM backgrounds channels; $ZZ$, $t\bar{t}Z$ and $WWZ$, respectively. We have used similar color conventions for SM backgrounds as in Fig.~\ref{fig:lep_meff14} and \texttt{BP1, BP4} and \texttt{BP5} are portrayed by yellow, cyan, and red regions respectively. Similar to 14 TeV analysis, it is evident that a large cut on $p_T^{l_1}$ and $m_{eff}$ will effectively discard SM backgrounds while merely affecting the signals. We obtain that a cut-set consisting of $N_l\geq4 + Z$ veto + 
$p_T^{l_1} > 150$ GeV  +$b$ veto along with strong $m_{eff}$ cut provides the maximal signal significance.  We define two signal regions, \texttt{SR-C} with $m_{eff}>1500$ GeV and \texttt{SR-D} with $m_{eff}>2200$ GeV. The cut flow table for selected benchmark points and the total SM backgrounds are presented in the Table.~\ref{tab:cut_flow_27} along with signal significance without and with $5\%$ systematic uncertainty. The signal significance for $\sigma_{ss}^0$ ($\sigma_{ss}^5$) for \texttt{BP1}, \texttt{BP4} and \texttt{BP5} are 22.77 (14.93), 9.04 (8.14), 2.35 (2.31) for \texttt{SR-D}.

The projected discovery ($5\sigma$) and exclusion ($2\sigma$) reach at the HE-LHC using cut-and-count based method are presented in Fig.~\ref{fig:reach_27tev_ml} in the slepton-LSP mass plane. The light-green shaded region denotes $2\sigma$ reach and $5\sigma$ reach is denoted by the dotted green line. The current limit obtained from ATLAS collaboration from Run-II data is marked with red color. We have observed that slepton masses can be excluded up to 2.86 (2.82) TeV with the choice of $\mlspone = 1.9$ (2.81) TeV with $95\%$ C.L.. The  $5\sigma$ projection reaches up to $\sim$ 2.32 TeV for $\mslep$ at the HE-LHC with cut-based analysis.

%%%%%%%%%%%%%%%%%%%%%%%%%%%%%%%%%%%%%
\begin{figure}[h]
\centering
\includegraphics[width=0.45\linewidth]{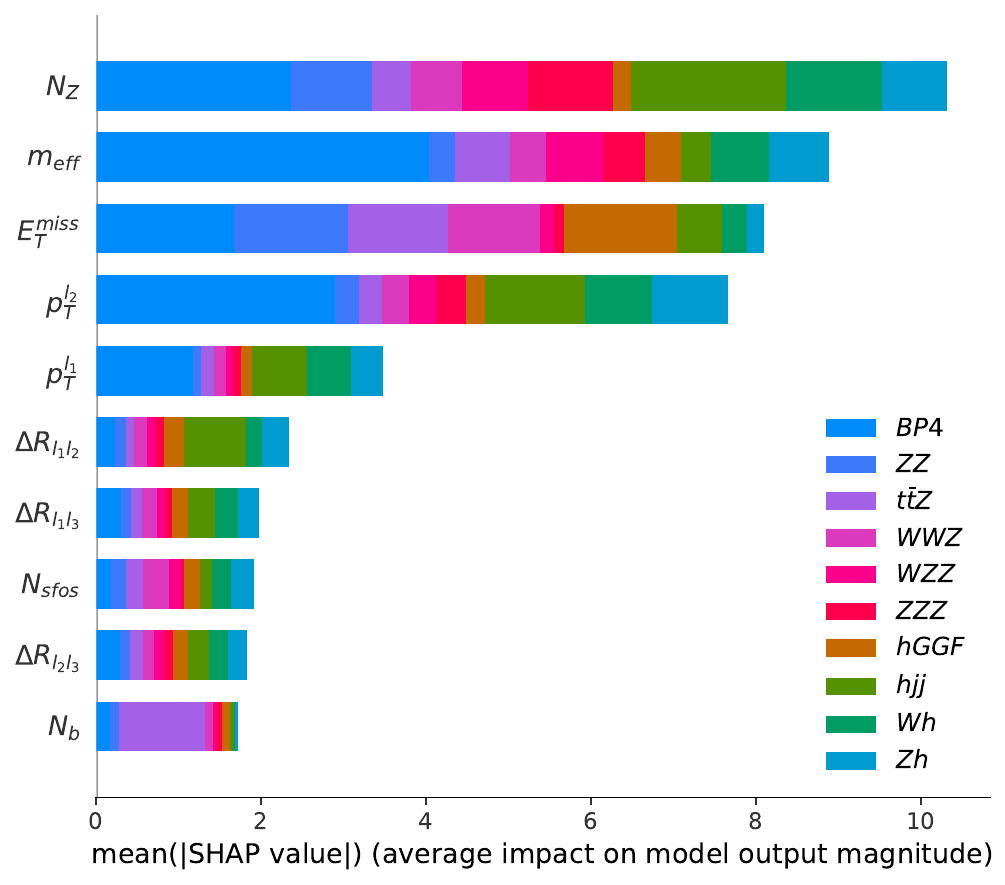}
\includegraphics[width=0.45\linewidth]{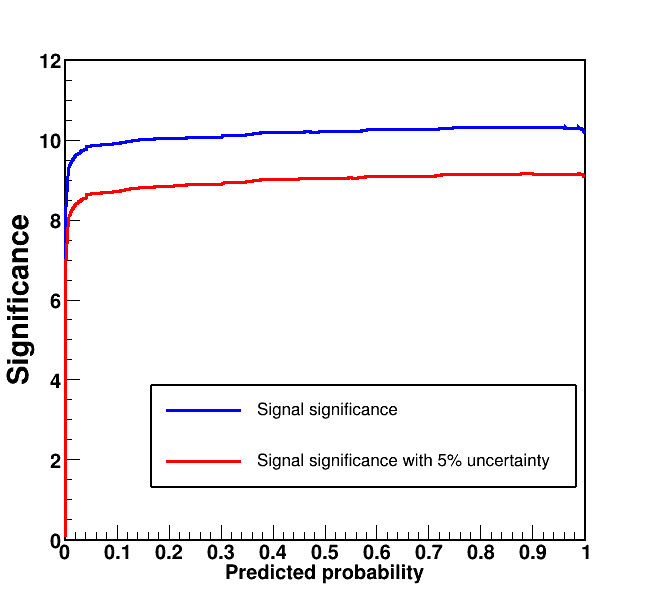}
\caption{Left: Shapley feature importance plot for the top ten variables for benchmark point BP4 and the SM backgrounds. Right: The signal significance at the HE-LHC without any systematic uncertainty and with 5\% uncertainty are displayed via blue and red lines as a function of predicted probability. }
\label{fig:shap_scoreBP4}
\end{figure}
%%%%%%%%%%%%%%%%%%%%%%%%%%%%%%%%%%%%%

%%%%%%%%%%%%%%%%%%%%%%%%%%%%%%%%%%%%%%%
\begin{table}[!htb]
%\small
\begin{tabular}{|c|c|c|c|c|c|}
\hline

Benchmark Points & Probability Score& Signal & Total   & Signal  &  Gain in $\sigma_{ss}$ \\
 &  &  & Background Yield &  Significance $\sigma_{ss}$ & from  \\
& &  &  &  (Sys Unc. = $5\%$) & Cut-based \\
\hline
\hline
\texttt{BP1} & 0.90  & 1098.85 & 10.73 & 32.98 (16.98) & 18\% (6\%)\\
%\cline{2-6}
 & 0.96 & 1080.31 & 4.18 & 32.80 (17.02)  & 17\% (6\%)\\
\hline
\hline
\texttt{BP4} & 0.90  & 107.68  & 1.23 & 10.31 (9.14) &  14\% (12\%)\\
%\cline{2-6}
 & 0.96 & 107.40 & 0.48 & 10.34 (9.18) & 14\% (12\%)\\
\hline
\hline
\texttt{BP5} & 0.90  & 9.92  & 0.21 & 3.11 (3.07) &  32\% (33\%) \\
%\cline{2-6}
 &  0.96 & 9.90  & 0.18 & 3.12 (3.08) & 33\% (33\%)\\
\hline
\hline
\end{tabular}
\caption{Signal yield, total background yield, the signal significance and the gain in significance at the HE-LHC using ML-based algorithm for different probability scores are shown here. In the last two columns, the numbers in the parenthesis correspond to signal significance and gain with systematic uncertainty $\epsilon = 5\%$.}
\label{tab:significance_27_ml}
\end{table}
%%%%%%%%%%%%%%%%%%%%%%%%%%%%%%%%%%%%%

%%%%%%%%%%%%%%%%%%%%%%%%%%%%%%%%%%%%
\begin{figure}[!htb]
\begin{center}
\scalebox{0.8}{\input{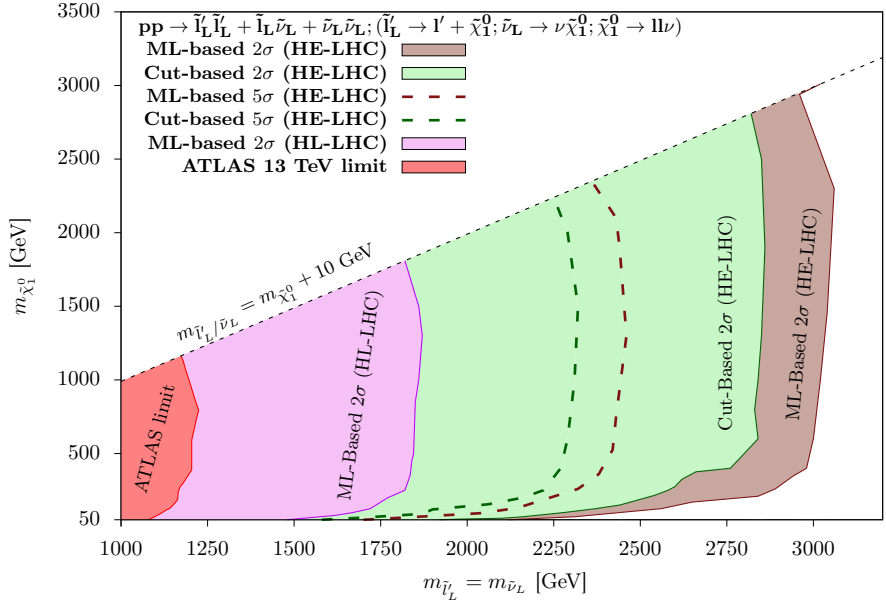}}
\caption {Projected discovery ($5\sigma$) and exclusion ($2\sigma$) reach in the slepton-LSP mass plane at the HE-LHC are presented for cut-based and ML-based analysis. The red regions represent the existing limit obtained by the ATLAS collaboration from Run-II data \cite{ATLAS:2021yyr}. The violet regions correspond to $2\sigma$ projected reach from the ML-based method at the HL-LHC as shown in Fig.\ref{fig:reach_14tev_ml}. The $5\sigma$ projected discovery reach obtained from traditional cut-based method and {\tt XGBoost} are shown via green and brown dotted lines respectively. The green and brown solid regions stand for $2\sigma$ projected to reach from cut-based and ML-based methods respectively at the HE-LHC.}

\label{fig:reach_27tev_ml}
\end{center}
\end{figure}
%%%%%%%%%%%%%%%%%%%%%%%%%%%%%%%%%%%

We proceed with a similar ML analysis as in Sec.\ref{sec:14_cut} to ascertain how much the sensitivity can be improved upon. We choose the same set of features and hyperparameter selection as discussed earlier. Similar to HL-LHC analysis, we obtain a similar Shapley feature importance plot which is presented in Fig.~\ref{fig:shap_scoreBP4}. On the right panel of Fig.~\ref{fig:shap_scoreBP4}  we present the variation of signal significance for the HE-LHC as a function of probability score. Next, we estimate the signal yields, total background yield, and signal significance for different values of probability score and systematic uncertainty for the 
three benchmark points. We show these results in the Table~\ref{tab:significance_27_ml}. In the last column, we present the gain in 
$\sigma_{ss}$. An improvement of $\sim$ 14\% to 33\% is observed in the {\tt XGBoost} results compared to the conventional cut-based method. 
 The numbers in the parentheses represent the results corresponding to 5\% systematic uncertainty. The improvement of signal significance is also echoed in the exclusion plot illustrated in the slepton-LSP mass plane (Fig.~\ref{fig:reach_27tev_ml}). The brown dotted line corresponds to the $5\sigma$ projections at the HE-LHC using ML based algorithm and it reaches up to $\sim$ 2.46 TeV. The 
 green dotted line (corresponding to cut-based $5\sigma$) indicates that the reach is enhanced by 140 GeV in the ML-based analysis. We also observe that the projected exclusion curve (displayed via brown solid regions) reaches up to $\sim$ 3.06 TeV resulting an improvement of $\sim$ 200 GeV as compared to the cut-based method (green region).
 
\subsection{Comparision among different scenarios}

%%%%%%%%%%%%%%%%%%%%%%%%%%%%%%%%%%%%%%%%
\begin{table}[!htb]
%\begin{center}
%\centering
\begin{tabular}{||c|c|c|c|c|c|c|c|c|c||} 
    \hline
    \multicolumn{5}{||c|} {HL-LHC} & \multicolumn{5}{c||} {HE-LHC} \\
    \hline
	Benchmark & \multicolumn{4}{c|}{\underline{~~~Signal Significance~~}} & Benchmark & \multicolumn{4}{c||}{\underline{~~~Signal Significance~~}}\\
	points & S-I & S-II & S-III & S-IV &points & S-I & S-II & S-III & S-IV \\
	\hline
    \texttt{BP1} & 8.35 & 6.69 & 4.84 & 3.52 & \texttt{BP1} & 32.98 & 26.64 & 18.86 & 13.50 \\
    \hline
    \texttt{BP2} & 6.39 & 5.25 & 3.86 & 2.83 & \texttt{BP4} & 10.31 & 8.46 & 6.26 & 4.59 \\
    \hline
    \texttt{BP3} & 2.40 & 1.91 & 1.32 & 0.93 & \texttt{BP5} & 3.11 & 2.51 & 1.76 & 1.18 \\
    \hline\hline
$m_{\lspone}$ [GeV] & \multicolumn{4}{c|}{Exclusion limit on $m_{\tilde{l}^{\prime}} = m_{\tilde{\nu}_L}$ [GeV]}  & $m_{\lspone}$ [GeV] & \multicolumn{4}{c||}{Exclusion limit on $m_{\tilde{l}^{\prime}} = m_{\tilde{\nu}_L}$ [GeV]} \\
\hline
800 & 1850 & 1800 & 1670 & 1575 & 1000 & 3020 & 2880 & 2620 & 2500 \\
\hline\hline
\end{tabular}
\caption{Comparison of signal significance of signal benchmark points at the HL-LHC and HE-LHC for different model scenarios using ML based analysis. }
\label{tab:scenario_compare}
\end{table}
%%%%%%%%%%%%%%%%%%%%%%%%%%%%%%%%%%%%%%% 

In this work, we have focused on the RPV SUSY scenarios with non-zero values of $\lambda_{121}$ and/or $\lambda_{122}$ which provide maximum leptonic ($l\equiv e,~\mu$) branching ratios of the LSP decay and thus leads to the most stringent limits on the slepton masses. Before concluding we comment on the collider limits in other scenarios associated with the remaining seven non-zero $\lambda_{ijk}$ couplings from $N_l \geq 4$ ($l \equiv e, \mu$) final state. The scenarios with non-zero $\lambda_{121}$ and/or $\lambda_{122}$, for which we already obtained the limits in Sec.~\ref{sec:14_cut} and Sec.~\ref{sec:27_cut} are denoted as {\textbf{S-I}}. 
\begin{itemize}
\item \underline{\textit{Scenario-II}}: \textbf{S-II} represents the models  
with single non zero couplings $\lambda_{i3k}$ where $i \equiv 1,~2$ and $k \equiv 1,~2$. The LSP pair in the final state decays to $4l$, $3l1\tau$, and $2l2\tau$ final states with 25\%, 50\%, and 25\% branching ratios respectively.  
\item  \underline{\textit{Scenario-III}}: \textbf{S-III} defines the model with non-zero value of $\lambda_{123}$ coupling where we get $2l2\tau$ final state with 100\% branching ratio from the LSP pair.  
\item  \underline{\textit{Scenario-IV}}:  \textbf{S-IV} corresponds to the non-zero values of $\lambda_{i33}$ with $i \equiv 1,~2$ where the LSP pair decays to 
 $2l2\tau$, $1l3\tau$ and $4\tau$ final states with 25\%, 50\% and 25\% branching ratios respectively. 
 \item Among these four scenarios, the $N_l \geq 4$ ($l = e, \mu$) channel provides the best limit in {\textbf{S-I}} models. On the other hand, one expects the weakest limit in \textbf{S-IV} model which is 
 mostly tau enriched. Discovery and exclusion reach in other two models (\textbf{S-III} and \textbf{S-IV}) or models with any arbitrary combination of non-zero   $\lambda_{ijk}$ lie between  \textbf{S-I} and \textbf{S-IV}. This pattern is observed in the Table~\ref{tab:scenario_compare} where we present 
the signal significance obtained from ML-based analysis at the HL-LHC and HE-LHC for the above mentioned four models. 
\item In Table~\ref{tab:scenario_compare}, we also show the variation of projected 
2$\sigma$ limits on slepton mass for a fixed choice of LSP mass. We observe that the limits are weaker by 275 (520) GeV at the Hl-LHC and HE-LHC 
in the extreme tau enriched model i.e., \textbf{S-IV} compared to \textbf{S-I}.
%\tcm{ \item We also consider the LR models with mass degenerate first two generation left and right handed slepton and benchmark studies show that limits are similar/weaker/stronger ?}
\end{itemize}

%%%%%%%%%%%%%%%%%%%%%%%
\section{Conclusion}
\label{sec:conclusion}
%%%%%%%%%%%%%%%%%%%%%%%
Supersymmetric signals have been searched extensively at the LHC. The R-parity conserving framework has typically been more widely studied compared to its R-parity violating counterpart. The large number of final states non-zero R-parity violating couplings can open up a need to be studied to ascertain the full extent of impact the LHC data has or is going to have on the relevant parameter space. Sleptons are of particular importance in the context of any SUSY scenarios since they affect the contribution of the model to some very crucial observables, such as light neutrino mass and mixing, lepton magnetic moment, and lepton number or lepton flavor violating decay rates. Non-zero $\lambda$ couplings also contribute to these observables. In this work, therefore, we have explored a scenario where the sleptons and sneutrinos are produced through usual RPC couplings and subsequently decay into leptons and neutrinos through RPV decay of the bino LSP. In order to understand the maximum impact LHC can have on the parameter space we consider the production of all three generations of sneutrinos and left-handed sleptons which have larger cross-sections compared to their right-handed counterpart. We have performed a detailed cut-based collider analysis alongside using a machine learning algorithm for comparison. Since we are interested in electron and muon-enriched final states, we have only considered non-zero $\lambda_{121}$ and $\lambda_{122}$ couplings. Our final state has rich lepton multiplicity, $N_{l} \geq$ 4 ($l \equiv e, \mu$). We observe that this final state can probe slepton and sneutrino masses most efficiently. In our analysis, we have assumed the left-handed sleptons and sneutrinos to be mass degenerate for simplicity. We explore both the high luminosity and high energy options of the LHC and present our results in terms of exclusion region ($2\sigma$ statistical significance) and discovery reach ($5\sigma$ statistical significance) in the LSP - slepton/sneutrino mass plane. Our study reveals that 
the discovery regions reach up to $\sim$ 1.49 TeV and $\sim$ 1.54 TeV while the exclusion regions reach up to $\sim$ 1.75 TeV and $\sim$ 1.87 TeV at HL-LHC for cut-based and ML algorithm respectively. Similarly at HE-LHC, we have shown that the projected discovery regions reach up to $\sim$ 2.32 TeV and $\sim$ 2.46 TeV while projected exclusion limits are $\sim$ 2.86 TeV and $\sim$ 3.06 TeV for cut-based and ML algorithm respectively. We obtain overall better improvement in the case of HE-LHC with ML algorithm. Understandably, the presence of other non-zero $\lambda$ couplings can give rise to $\tau$-leptons in the final states and as a result reduces the signal efficiency. We have compared the different possible scenarios with different non-zero $\lambda$ couplings. We observed that in the worst case scenario when only the $\lambda_{i33}~(i=1,2)$ are non-zero, the signal significance drops by a factor of $\sim$ 2.2$-$2.6  at HL-LHC and HE-LHC. 

\vspace{+1cm}

\noindent
\textbf{Data availability statement:} No data associated in the manuscript.

\bibliography{sn-bibliography}

\end{document}